\documentstyle[12pt]{article}

\def\be{\begin{equation}}
\def\ee{\end{equation}}
\def\bea{\begin{eqnarray}}    
\def\eea{\end{eqnarray}}

\newcommand{\na}{\nabla}  
\newcommand{\al}{\alpha}  
\newcommand{\ga}{\gamma}  
\newcommand{\de}{\delta}  

\begin{document}
\begin{flushright}
\end{flushright}
\pagestyle{plain}
\begin{center}
\LARGE{\bf Conformal Invariance and Quantum Aspects of Matter\\}
\vspace{.5cm}
\small
\vspace{1.5cm}
{\Large{\bf H. Motavali$^{1}$}}\footnote{e-mail address: m7412926@cic.aku.ac.ir},~~
{\Large{\bf H. Salehi$^{2,3}$}}\footnote{e-mail address: h-salehi@cc.sbu.ac.ir}
and {\Large{\bf M. Golshani$^{4}$}}\footnote{e-mail address: golshani@ihcs.ac.ir}  \\  \vspace{0.5cm}
\small{$^{1}$ Department of Physics, Amirkabir University of Technology, 15875-4413, Tehran, Iran \\
$^{2}$ Department of Physics, Shahid Beheshti University, Evin, Tehran 19834,  Iran \\ 
$^{3}$ Institute for Humanities and  Cultural Studies, 14155-6419, Tehran, Iran\\
$^{4}$ Department of Physics, Sharif University of Technology, 11365-9161, Tehran, Iran}\\
\today 
\end{center}
\vspace{.5cm}
\small
\begin{abstract}
The vacuum sector of the Brans-Dicke theory is studied from the viewpoint of a non-conformally
invariant gravitational model. We show that, this theory can be conformally 
symmetrized using an appropriate conformal transformation. The resulting 
theory allows a particle interpretation, and suggests that the quantum 
aspects of matter may be geometrized.
\end{abstract}
\section{Introduction}

Conformal invariance is one of the desirable symmetries in modern theories of physics. 
The principle of conformal invariance states that all of the fundamental equations of physics
should be fully invariant under local transformations of the units of length
and time ${}^1$. In principle, such transformations correspond to stretching all lengths 
and durations by the space-time dependent conversion factors $\Omega$.
Conformal transformation techniques have been widely used in scalar-tensor theories
as alternative theories of gravity.

In this paper, we shall show that these techniques can be used to conformally 
symmetrize a non-conformally invariant 
theory. The resulting conformally invariant theory may be interpreted from 
different viewpoints with respect to the first one. As a model, we shall take   
the non-conformally invariant Brans-Dicke theory and show that this theory 
transforms to a conformally invariant one for which a particle 
interpretation can be derived. We shall use units in which $\hbar$=c=1.

\section {Duality Transformation}
We start with the vacuum part of the Brans-Dicke theory described by the action
\begin{eqnarray}
S[\Phi]=\frac{1}{2k}{ \int d^4x {\sqrt {-g}} \{ \Phi R- \frac{\ln {\bf \beta}}{\Phi} g^{\mu\nu} \na_\mu \Phi \na_\nu\Phi \}}
\end{eqnarray}
where $k=8 \pi G$,$~~ R$ is the curvature scalar, $\Phi$ denotes a real scalar field nonminimally coupled to the gravity, 
and $\ln {\bf \beta}$ represents Brans-Dicke parameter.
One can easily check that the action (1) is not invariant under the conformal transformation 
\begin{eqnarray}
 g_{\mu \nu} \longrightarrow \Omega^2(x) g_{\mu\nu}  ~~~,~~~\Phi \longrightarrow \Omega^{-2}(x)\Phi  \nonumber
\end{eqnarray}
unless $\ln{\bf \beta}=-3/2$. This action is invariant under what we call the duality transformation
\begin{eqnarray}
(\Phi)  \longrightarrow  (\Phi)^D
\end{eqnarray}
where we have used the abbreviations
\begin{eqnarray}
(\Phi) = (g_{\mu \nu},~\Phi ,~ {\bf \beta} )~~~~~and~~~~~(\Phi)^D = (\Omega^2 g_{\mu \nu},~\Omega^{-2}\Phi ,~ {\bf \beta}^{-1}) 
\nonumber
\end{eqnarray}
with $\Omega=(G\Phi)^\alpha$ in which $\alpha =\frac{1}{2}[1 \pm \sqrt {\frac{3-2\ln {\bf \beta}}{3+2\ln {\bf \beta}}}]$. 
The duality transformation (2) is a special case of general transformations introduced in Ref. 2.
This transformation tells us that, in the vacuum sector there is no
distinction between large and small values of ${\bf \beta}$, or equivalently between positive 
and negative values of Brans-Dicke parameter $\ln{\bf \beta}$. Thus, one can not estimate
this parameter definitely, unless matter field is present. In principle, there is no distinction 
between $(\Phi)$ and its dual $(\Phi)^D$, and one may take these as equivalent configurations in the vacuum.
However, a distinction between $(\Phi)$ and $(\Phi)^D$ can be made if the 
sign of the Brans-Dicke parameter $\ln{\bf \beta}$ is important. Typically, if the 
action (1) is coupled to the matter field, the observational constraint implies $\ln{\bf \beta} > 500$. 
While in the model which we present in a later section, particle interpretation 
requires $\ln{\bf \beta} < -\frac{3}{2}$. 
\section {The Conformally Transformed Action}
To conformally symmetrize the action (1), first we put (1), with redefinitions 
$\Phi=G\phi^2$ and $\ln{\bf \beta}=-\frac{3}{2}(1+\xi)$, into the form
\begin{eqnarray}
S[\phi]={ \frac{3}{8\pi}{ \int  d^4x  \sqrt{-g} \{ 
         \frac{1}{6} {\phi}^2 R+(1+\xi) g^{\mu \nu} \na_\mu \phi \na_\nu\phi+ \frac{1}{2} \lambda \phi^4   \} }} 
\end{eqnarray}
where $\frac{1}{2} \lambda \phi^4$ is a cosmological function 
which is added for later usage, 
$\lambda$ being a dimensionless parameter.
This action is not invariant under the  conformal transformation 
\begin{eqnarray}
g_{\mu \nu} \longrightarrow \Omega^2(x) g_{\mu\nu}~~~~,~~~~ \phi \longrightarrow \Omega^{-1}(x) \phi
\end{eqnarray}
unless $\xi=0$. 
Thus, one may consider $\xi$ as the deformation parameter which characterizes 
the conformal invariance breaking. 

Applying the conformal transformation 
(4) to the action (3) one sees that all terms are invariant, except the term 
$\sqrt{-g} \xi g^{\mu \nu} \na_\mu \phi \na_\nu\phi $ which 
transforms as
\begin{eqnarray}
\sqrt{-g} \xi g^{\mu \nu} \na_\mu \phi \na_\nu\phi \longrightarrow \sqrt{-g} \phi^2 g^{\mu \nu} \na_\mu S \na_\nu S
\nonumber
\end{eqnarray}
where we have used the definition 
\begin{eqnarray}
S=\sqrt{\xi} \ln(\phi l/\Omega)
\end{eqnarray}
in which $l$ is a fundamental length. The introduction of this length is necessary to make the argument 
of the logarithm dimensionless. The point now is that $S$ as a dimensionless field 
does not transform under a subsequent conformal transformation. \\
Therefore, applying the conformal transformation (4) and using the definition 
(5), the action (3) transforms to
\begin{eqnarray}
S[\phi]={ \frac{3}{8\pi}{ \int  d^4x  \sqrt{-g} \{ g^{\mu \nu} \na_\mu \phi \na_\nu\phi 
        +( g^{\mu \nu} \na_\mu S \na_\nu S + \frac{1}{6} R) {\phi}^2 
       +\frac{1}{2} \lambda \phi^4 \} }}
\end{eqnarray}
which is exactly conformally invariant.

Variations of $S[\phi]$ with respect to  $\phi$ and $S$ lead, respectively, to  
\begin{eqnarray}
\Box^g \phi - (g^{\mu \nu} \na_\mu S \na_\nu S +\frac{1}{6}R )\phi- \lambda \phi^3  =0 
\end{eqnarray}
and
\begin{eqnarray}
\na_\mu (\phi^2 \na^\mu S)=0
\end{eqnarray}
where $\Box^g=g^{\mu\nu}\nabla_\mu\nabla_\nu$.

Variation of $S[\phi]$ with respect to $g_{\mu \nu}$ gives the 
gravitational equation, but we shall use here a non-variational technique
to derive this equation. For this purpose we multiply the field Eq. (7) 
by $\na_\mu \phi$
\begin{eqnarray}
\Box^g \phi\na_\mu \phi -(\na_\al S \na^\al S +\frac{1}{6}R)\phi \na_\mu \phi - \lambda \phi^3 \na_\mu \phi  =0. 
\nonumber
\end{eqnarray}
The first term can be rewritten as
\begin{eqnarray}
\Box^g \phi\na_\mu \phi =\na_\nu (\na_\mu \phi \na^\nu \phi 
-\frac{1}{2} \de_\mu^{~\nu} \na_\al \phi \na^\al \phi). 
\nonumber
\end{eqnarray}
The term $\phi \na_\mu \phi \na_\al S \na^\al S$ may be rewritten in the form
\begin{eqnarray}
\phi \na_\mu \phi \na_\al S \na^\al S &=&\frac{1}{2}(\na_\mu \phi^2) \na_\al S \na^\al S  \nonumber \\
                                      &=&\frac{1}{2} \na_\nu( \de_\mu^{~\nu} \phi^2 \na_\al S \na^\al S -2 \phi^2 \na_\mu S \na^\nu S) \nonumber
\end{eqnarray}
where we have used Eq. (8). \\
Taking into account the Einstein tensor $G_{\mu\nu}=R_{\mu\nu}-\frac{1}{2}g_{\mu\nu}R$, one finds 
\begin{eqnarray}
\de_\mu^{~\nu}R \phi \na_\nu \phi &=&R_\mu^{~\nu} \na_\nu \phi^2 - \na_\nu( G_\mu^{~\nu}\phi^2) \nonumber \\
                    &=&\na_\nu[(\na_\mu \na^\nu -\de_\mu^{~\nu} \Box^g-G_\mu^{~\nu})\phi^2] \nonumber 
\end{eqnarray}
where we have used the definition of the curvature tensor
\begin{eqnarray}
[ \na_\al , \na_\beta ] V_\ga=-R^\rho_{~\ga\al\beta} V_\rho \nonumber
\end{eqnarray}
in which $V_\ga$ is some covariant vector field.\\
Finally, the last term can be rewritten as  
\begin{eqnarray}
\lambda \phi^3 \na_\mu \phi=\frac{1}{4} \na_\nu(\de_\mu^{~\nu} \lambda \phi^4). \nonumber
\end{eqnarray}

Combining all these relations one gets
\begin{eqnarray}
\na^\nu T_{\mu \nu}=0
\end{eqnarray}  
where $T_{\mu \nu}=T_{\mu \nu}^\phi+T_{\mu \nu}^S$, in which 
the first term  
\begin{eqnarray}
T_{\mu \nu}^\phi= \phi^2 G_{\mu \nu}+(g_{\mu \nu} \Box^g - \na_\mu \na_\nu)\phi^2 + 6\na_\mu \phi \na_\nu \phi  
-3g_{\mu \nu}\na_\alpha \phi \na^\alpha \phi - \frac{3}{2}\lambda \phi^4 g_{\mu \nu}
\nonumber
\end{eqnarray}
is the so-called conformal stress tensor${}^3$, and the second term is
\begin{eqnarray}
T_{\mu \nu}^S= 6\phi^2 ( \na_\mu S \na_\nu S -\frac{1}{2} g_{\mu \nu} \na_\alpha S \na^\alpha S).
\nonumber
\end{eqnarray}
If the scalar field $\phi$ has a constant configuration, then 
$\na^\nu T_{\mu \nu}^\phi=0$, and so $T_{\mu \nu}^S$ must be conserved in Eq. (9).
A varying configuration of $\phi$ leads, in general, to non conserved  
tensor $T_{\mu \nu}^S$.  \\
Now, if we want to couple the action (6) to the free gravitational action, we obtain 
\begin{eqnarray} 
G_{\mu\nu}=-kT_{\mu\nu}.
\nonumber
\end{eqnarray}
Since the conformal invariant action (6) has no free 
gravitational part in the vacuum sector, we have $T_{\mu\nu}=0$. Thus, 
one finds
\begin{eqnarray}
\phi^2 G_{\mu \nu}+(g_{\mu \nu} \Box^g - \na_\mu \na_\nu)\phi^2 + 6(\na_\mu \phi \na_\nu \phi+ \phi^2 \na_\mu S \na_\nu S)  \\ \nonumber
-3g_{\mu \nu}(\na_\alpha \phi \na^\alpha \phi + \phi^2 \na_\alpha S \na^\alpha S) -\frac{3}{2}\lambda \phi^4 g_{\mu\nu}=0
\end{eqnarray}
which is the gravitational equation.
This equation can be derived in the standard way from the variation of the action (6) 
with respect to $g_{\mu \nu}$.
\section {Particle Interpretation}   
The action (6) and the corresponding Eqs. (7), (8) and (10) are exactly conformally invariant 
under the conformal transformation (4). This implies that the non-conformally invariant 
theory (3) which is introduced with symmetry breaking parameter $\xi$, 
may be symmetrized using the conformal transformation (4) and definition (5). This
definition constraints $\xi$ to positive values, or equivalently 
$\ln{\bf \beta} <-\frac{3}{2}$.

Now, we study a particular way to arrive at a particle interpretation of the 
theory given by the action (6).
Let us rewrite Eq. (7) as 
\begin{eqnarray}
g^{\mu \nu} \na_\mu S \na_\nu S=-\lambda \phi^2 +\frac{ \Box^g \phi}{\phi}-\frac{1}{6}R.
\end{eqnarray}
Conformal invariance of this equation, or equivalently conformal symmetry of the 
action (6) implies that various frames can be assigned 
to the theory given by this action, depending on the particular configuration which
one chooses for the scalar field $\phi$.  

We first study a frame by the condition
$\phi= \phi_0$ = const. 
In this frame Eqs. (8) and (11) take, respectively, the forms
\begin{eqnarray}
\Box^g S=0
\end{eqnarray}
and
\begin{eqnarray}
g^{\mu \nu} \na_\mu S \na_\nu S=- {M_0}^2 -\frac{1}{6}R
\end{eqnarray}
where $M_0^2=\lambda \phi_0^2$.

In the limit $g_{\mu\nu} \longrightarrow \eta_{\mu\nu}$, in which $\eta_{\mu\nu}$ 
is the Minkowski metric, the last equation takes the 
form of Hamilton-Jacobi equation characterizing an ensemble of relativistic 
particle. At every point a particle has a four-momentum $\na_\mu S$, and 
mass $M_0$. We may call this frame the classical frame. 

To define another frame, we can apply the conformal transformation  
\begin{eqnarray}
g_{\mu \nu} \longrightarrow \Omega^2(x) g_{\mu\nu}~~~~,~~~~ \phi_0 \longrightarrow \Omega^{-1}(x) \phi_0
\end{eqnarray}
to the classical frame. 
In the new frame, Eq. (13) takes the form
\begin{eqnarray}
g^{\mu \nu} \na_\mu S \na_\nu S=-{M_0}^2 + \frac{ \Box^g \Omega}{\Omega}-\frac{1}{6}R
\end{eqnarray}
where we have used the transformation law of the Ricci curvature $R$ under (14) 
\begin{eqnarray}
R \longrightarrow \Omega^{-2} ( R-6\frac{\Box^g \Omega}{\Omega}).
\nonumber
\end{eqnarray}
Comparison of Eqs. (13) and (15) indicates that, the new frame is equivalent
to the classical frame, except for the particle mass which is modified by an extra term 
$\frac{ \Box^g \Omega}{\Omega}$. The physical interpretation of the extra term 
is postponed to the next section.    
      
In the new frame, Eq. (12) does not change, however, it is convenient to rewrite 
it in the following form 
\begin{eqnarray}
\na_\mu (\Omega^2 \na^\mu S)=M \frac{d}{d \tau}\Omega^2
\end{eqnarray}
where we have used
\begin{eqnarray}
\na_\mu S=M U_\mu  ~~~~~~~~~and~~~~~~~~~~   U_\mu \nabla^\mu =\frac{d}{d \tau}
\nonumber
\end{eqnarray}
in which $\tau$ is a parameter along the particle trajectory and $U_\mu$ is the
four-velocity of the particle.

\section{Relativistic Motion of the Particle}

In the new frame, the equation of motion of a particle in the ensemble may be  
described in terms of a pilot wave
\begin{eqnarray}
\psi= \Omega~e^{iS}.
\nonumber
\end{eqnarray}
This wave satisfies the equation 
\begin{eqnarray}
\Box^g \psi- {M_0}^2 \psi=(\frac{1}{6}R+2i M \frac{d}{d \tau} \ln |\psi|)\psi
\nonumber
\end{eqnarray}
which is obtained from the combination of Eqs. (15) and (16).

Under the assumption that the amplitude $\Omega$ is independent of the parameter 
$\tau$ along the particle trajectory, we get
\begin{eqnarray}
\Box^g \psi-M_0^2 \psi=\frac{1}{6}R\psi
\nonumber
\end{eqnarray}
which in the background approximation $g_{\mu\nu} \longrightarrow \eta_{\mu \nu}$ 
leads to the Klein-Gordon equation with the mass $M_0$.

The merit of introducing the wave $\psi$ is that it acts
as a sort of a pilot wave in the sense of causal interpretation of
quantum mechanics,${}^{4,5}$ the term $\frac{\Box^g \Omega}{\Omega} $ on the right
hand side of Eq. (15) being the associated quantum potential.
Therefore, we may call the new frame, the quantum one. 
The appearance of quantum potential is a consequence of transition from the classical 
frame to the quantum one by the application of the conformal transformation. 
This supports the idea that quantum effects may be contained in the conformal degree of freedom of the space-time
metric, and that the quantum aspects of matter may be geometrized.${}^{6,7}$

\end{document}